%% file: main.tex
\documentclass[
 amsmath,amssymb,
reprint,%
,floatfix]{article}

\usepackage{graphicx}
\usepackage{dcolumn}
\usepackage{bm}

\usepackage[utf8]{inputenc}
\usepackage[T1]{fontenc}
\usepackage{tabularx}
\usepackage{mathptmx}

\usepackage[colorlinks,allcolors=cyan!70!black]{hyperref} 
\usepackage[bottom]{footmisc}  
\usepackage{chemformula}   
\usepackage{float} 
\usepackage[noblocks]{authblk}
\usepackage[a4paper, total={6in, 8in}]{geometry}

\title{Orientation before destruction. A multiscale molecular dynamics study}

\date{}
\author[1]{Anna Sinelnikova}
\author[1,2]{Thomas Mandl}
\author[1]{Harald Agélli}
\author[1]{Oscar Grånäs}
\author[3]{Erik G. Marklund}
\author[1,4]{Carl Caleman}
\author[1,3]{Emiliano De Santis\thanks{emiliano.desantis@physics.uu.se}}

\affil[1]{Department of Physics and Astronomy, Uppsala University, Box 516, SE-751 20 Uppsala, Sweden}
\affil[2]{University of Applied Sciences Technikum Wien, H\"ochst\"adtplatz 6, A-1200 Wien, Austria}
\affil[3]{Department of Chemistry -- BMC, Uppsala University, Box 576, SE-751 23 Uppsala, Sweden}
\affil[4]{Center for Free-Electron Laser Science, DESY, Notkestrasse 85, DE-22607 Hamburg, Germany}

\begin{document}

    \maketitle

\begin{abstract}
The  emergence of ultra-fast X-ray free-electron lasers opens the possibility of imaging single molecules in the gas phase at atomic resolution.
The main disadvantage of this imaging technique is the unknown orientation of the sample exposed to the X-ray beam, making the three dimensional reconstruction not trivial. Induced orientation of molecules prior to X-ray exposure can be highly beneficial, as it significantly reduces the number of collected diffraction patterns whilst improving the quality of the reconstructed structure. We present here the possibility of protein orientation using a time-dependent external electric field. We used \textit{ab initio} simulations on Trp-cage protein to provide a qualitative estimation of the field strength required to break protein bonds, with 45 V/nm as a breaking point value.  Furthermore, we simulated, in  a classical molecular dynamics approach, the orientation of ubiquitin protein by exposing it to different time-dependent electric fields. The protein structure was preserved for all samples at the moment orientation was achieved, which we denote `orientation before destruction'. Moreover, we find that the minimal field strength required to induce orientation within ten ns of electric field exposure,  was of the order of 0.5 V/nm. 
Our results help explain the process of field orientation of proteins and can support the design of instruments for protein orientation.
\end{abstract}


\section{Introduction}
Proteins are essential components of all organisms, involved in a huge number of functions in living cells. It is generally accepted that a protein's function and three-dimensional (3D) structure are closely connected. Determining proteins tertiary structures is thus of crucial importance for basic science as well as for medicine. To date, one of the most important and widely used techniques for protein structure determination is X-ray crystallography. It relies on the coherent diffraction of X-rays from a crystallized sample of appropriate size and quality~\cite{bragg1913reflection}, which can not always be produced. Compounding this, proteins that are inherently dynamic might not crystallize in the relevant conformations.
The new era of X-ray Free-electron Lasers (XFELs), like  LCLS~\cite{emma2010first} or  European XFEL ~\cite{schneidmiller2011photon} facilities, able to produce a series of short and highly intense X-ray pulses, enabled a new technique for structure determination of isolated molecules in the gas phase -- Single Particle Imaging (SPI)~\cite{bogan2008single,neutze2000potential}. SPI overcomes the need of a crystalline sample, but the extreme intensity of the X-ray pulse causes severe radiation damage to the molecule, leading to coulombic explosion\cite{chapman2014diffraction}. Diffraction patterns are thus obtained from separate exposures of identical molecules, each in their own random orientation. Hence, a single diffraction pattern reflects a single view of the particle. The unknown orientation of the particle renders its 3D reconstruction problematic, complicated and expensive -- in terms of large sample consumption and beam-time required.
Huge  numbers of  diffraction patterns are needed in order for the reconstruction of the 3D structure of the molecule, which is accomplished using specific algorithms that have been developed to deal with this task~\cite{flamant2016expansion,ayyer2016dragonfly,schwander2012symmetries,mayne2013rapid,zhou2014multiple}. Data scarcity, sample heterogeneity, detector masking and background noise effects can often lead to artefacts on the reconstructed structures~\cite{lundholm2018considerations,hosseinizadeh2015single,poudyal2020single}.
We recently presented a potential solution in Marklund {\it et al.} study~\cite{Marklund2017a}, where classical molecular dynamics (MD) simulations were successfully used to show the possibility of orienting proteins by aligning their intrinsic dipole moments against an external, static, electric field (EF). We were able to define a range of the EF strengths which allow protein orientation without inducing significant structural loss, and also discovered that longer exposure times shifted the range toward lower field strengths.
Here, we present a natural continuation to the Marklund {\it et al.} study~\cite{Marklund2017a}, where we use MD to explore time-dependent EFs.
 We apply a multi-scale approach comprising both \textit{ab initio} and classical MD simulations. Using time-dependent density functional theory (TD-DFT) in the presence of an EF we identified at which field strengths covalent bonds remain intact in a protein, using the mini-protein Trp-cage as a model system. We moreover performed gas-phase classical MD simulations on ubiquitin to study the effect of the time-dependent EF on its orientation and the consequent structural evolution.
 Our results are of key for understanding the process of dipole orientation induced by an external, time-dependent EF. As such they can serve to guide in the design of an apparatus able to implement this phenomenon to manipulate proteins in SPI and other applications.

\section{Materials and Methods}

\subsection{\textit{Ab initio} MD simulations}
A molecule becomes polarized when subjected to external EFs as the charge in the molecule rearrange in order to screen the EF. Consequently, this generally changes the molecular dipole. Moreover, the rearrangements of the electronic structure result in residual forces on charged sites in the protein. We used the \textit{ab initio} MD software  package Siesta 4.1~\cite{Soler2002} to estimate the forces resulting from the interaction with the EF. We followed the same procedure as published in our earlier work~\cite{sinelnikova2020reproducibility}. \textit{Ab initio} calculations  were carried out on a small protein, Trp-cage (PDBid: 1L2Y)~\cite{TRP-Cage1L2Y}, with a total charge of +2 $e$, as expected in the vacuum state~\cite{patriksson2007protein}. We first thermalized the system using Born-Oppenhiemer MD, employing the Nosè thermostat to conserve a temperature of 300 K. For this step of the simulation, we used double-Z basis set with one polarization orbital per atom. The basis functions where generated with a shallow confinement potential of 0.001 Ry to allow for sufficient diffuse functions. The exchange-correlation integration grid was determined by a 200 Ry cut-off, and was treated according  to the Van der Waals function described by Vydrov and van Voorhis~\cite{VVvdw}. The thermalization simulation was two ps long, with a time step of 0.5 fs. Next, we oriented the protein  so that its dipole moment was aligned against the external EF. Without allowing for any nuclei dynamics  we then exposed  the protein to EFs ranging from 0.5 to 50.0 V/nm. For these simulations, we extended the basis set to encompass the charges in the electron distribution with respect to the ground state and used a triple-Z basis set with double polarization orbitals. For accurate partial charges the integration mesh cut-off was increased to 500 Ry. 
\subsection{Classical MD simulations} 
A set of gas-phase classical MD simulations were performed to study the orientation of ubiquitin exposed to a time-dependent EF. Gromacs 4.5.7~\cite{hess2008gromacs} simulation package was used together with the OPLS-AA force field~\cite{kaminski2001evaluation} in accordance with our previous studies of protein in gas phase~\cite{patriksson2007protein,marklund2009structural,sinelnikova2020reproducibility,mandl2020structural,Marklund2017a}.
In order to  sample sufficient statistics for our analysis and to better mimic the heterogeneity of gas-phase experimental samples, we performed independent sets of simulations starting from different protein structures. For this purpose, starting from coordinates  based on crystallographic data of ubiquitin~\cite{vijay1987structure} (PDBid: 1UBQ),  we ran a 10-ns pre-simulation in solution 
in the NPT ensemble with Berendsen~\cite{berendsen1984molecular} weak coupling. Temperature was set to 300 K with 0.1 ps time coupling and pressure was set to 1 bar with a time coupling of 20 ps. TIP4P~\cite{jorgensen1983comparison} water model was used. From the equilibrated portion of this bulk simulation (2.5 ns - 10 ns) we extracted structures at randomly picked times. 

\begin{figure}[hbt!]
\centering
\includegraphics[width=0.6\linewidth]{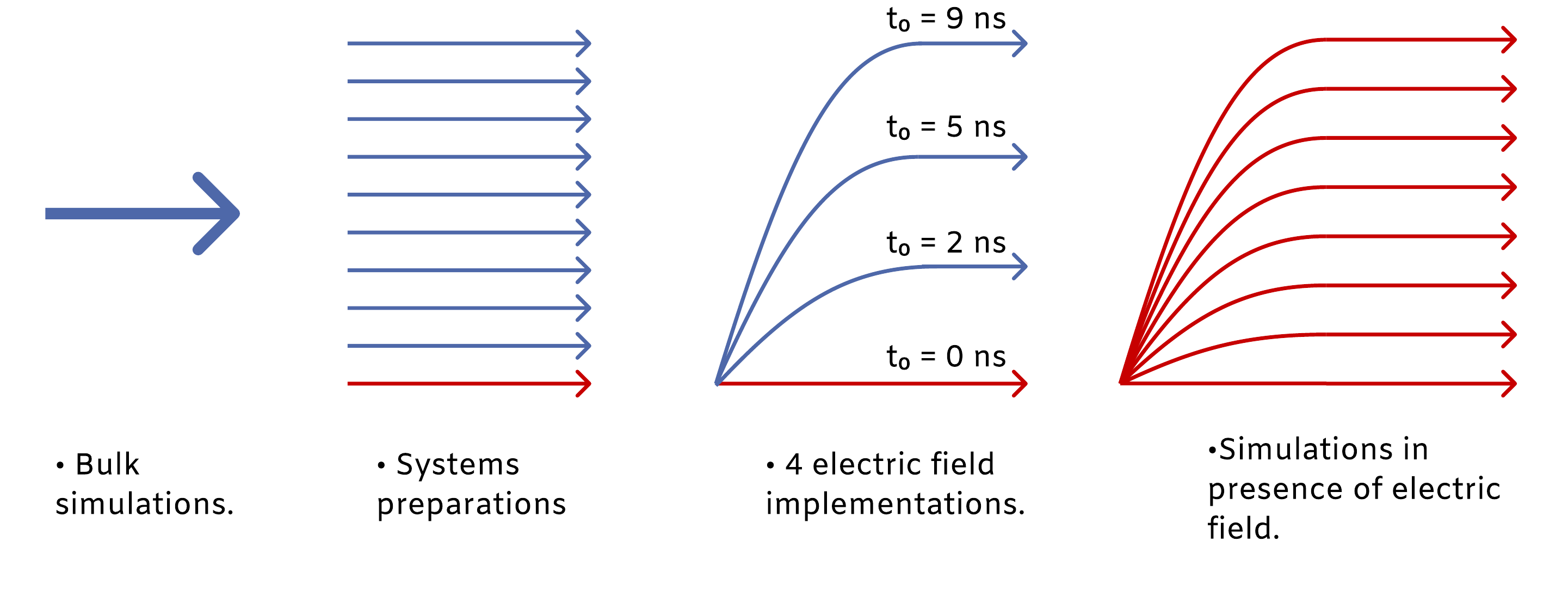}
\caption{\textit{Classical simulations} - The schematic representation of performed classical MD simulations. For more detailed description see Figure~\ref{fig:simScheme} of  Supporting  Information  section.}
\label{fig:sim_scheme_simple}
\end{figure}

The strategy we used then, depicted in Figures~\ref{fig:sim_scheme_simple} and ~\ref{fig:simScheme} of  Supporting Information section, is the following. After removing the solvent, we assigned the protonation states of ubiquitin in vacuum according to published data~\cite{breuker2002detailed,oh2002secondary,patriksson2007protein}, resulting in total charge of +7 $e$. The systems were relaxed in vacuum, and the temperature adjusted over 100 ps simulation to 300 K using the Berendsen thermostat~\cite{berendsen1984molecular}. We then ran a 10 ns long simulations in which we  allowed the structures  to equilibrate without thermostat. At the end of these pre-runs, the temperature of all the replicas was spanning a range of 305 K $\pm$ 5 K. Subsequently, we performed again 100 ps simulation with a temperature coupling at 300 K to ensure all the structures were at the same temperature. The structures obtained in this way were oriented to have their dipole moment parallel to the z-axis of the simulation box and were used as starting structures  to perform the EF orientation simulations. The time-dependent EF~\cite{caleman2008picosecond} was implemented as
\begin{equation}
 E(t)= E_{0} \exp{\frac{-(t-t_{0})^{2}}{2\sigma^2}} \cdot H(t_{0}-t) + E_{0}\cdot H(t-t_{0})
 \end{equation}
 where $H(t)$ is the Heaviside function, $\, t_{0}\in[0,2,5,9]$ ns and $ E_{0}\in [0.1, 0.2, 0.5, 0.8, 1.0, 1.5, 2.5, 3.0]$ V/nm. The direction of the EF was set to be parallel to the x-axis of the simulation  box.  Duration for simulations in which $t_{0}\in[0,2,5]$ ns was set to 10 ns and for $t_{0}=9$ ns  to 14 ns. In total, 320  independent simulations were performed (ten starting structures, four choices of $t_0$ value, eight EF strengths). \\

Long-range electrostatic forces in vacuum were captured using no cut-offs for non-bonded interactions. The equations of motion were propagated using the Leap-Frog integration scheme~\cite{van1988leap} with a 0.5 fs time step. To reproduce perfect vacuum, neither pressure coupling nor periodic boundary conditions were applied.

\section{Results}

\subsection {\textit{Ab initio} MD simulations: Validation of classical MD approach}
\begin{figure}[hbt!]
\centering
\includegraphics[width=0.7\linewidth]{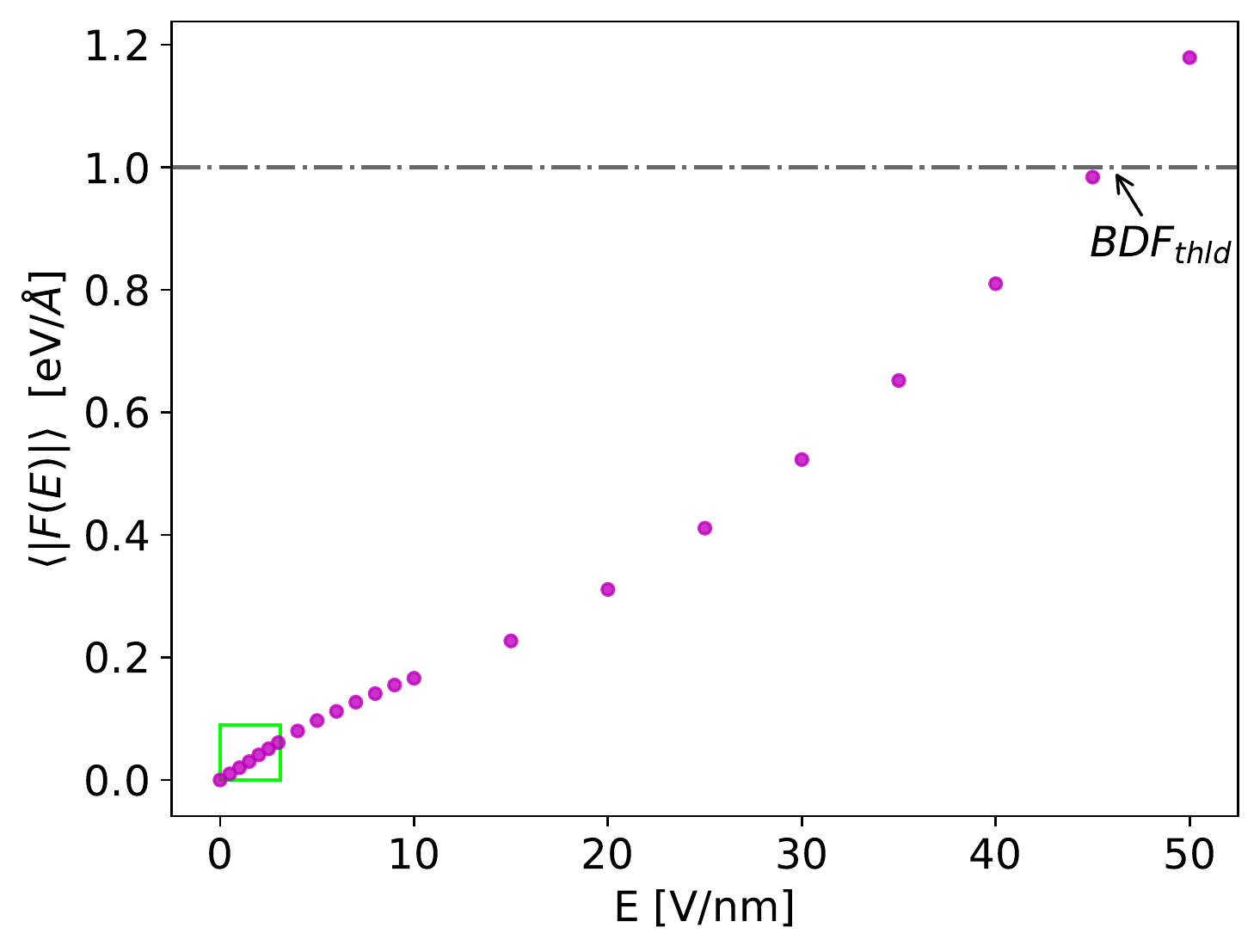}
\caption{\textit{Ab initio simulations} - Mean value of the force exerted on Tpr-cage atoms, defined in equation~\ref{eq:mean_force}, as a function of the electric field strength. The region of interest for the classical MD simulations is presented in the inset.}
\label{fig:average_force}
\end{figure}

\begin{figure}[hbt!]
\centering
\includegraphics[width=0.5\linewidth]{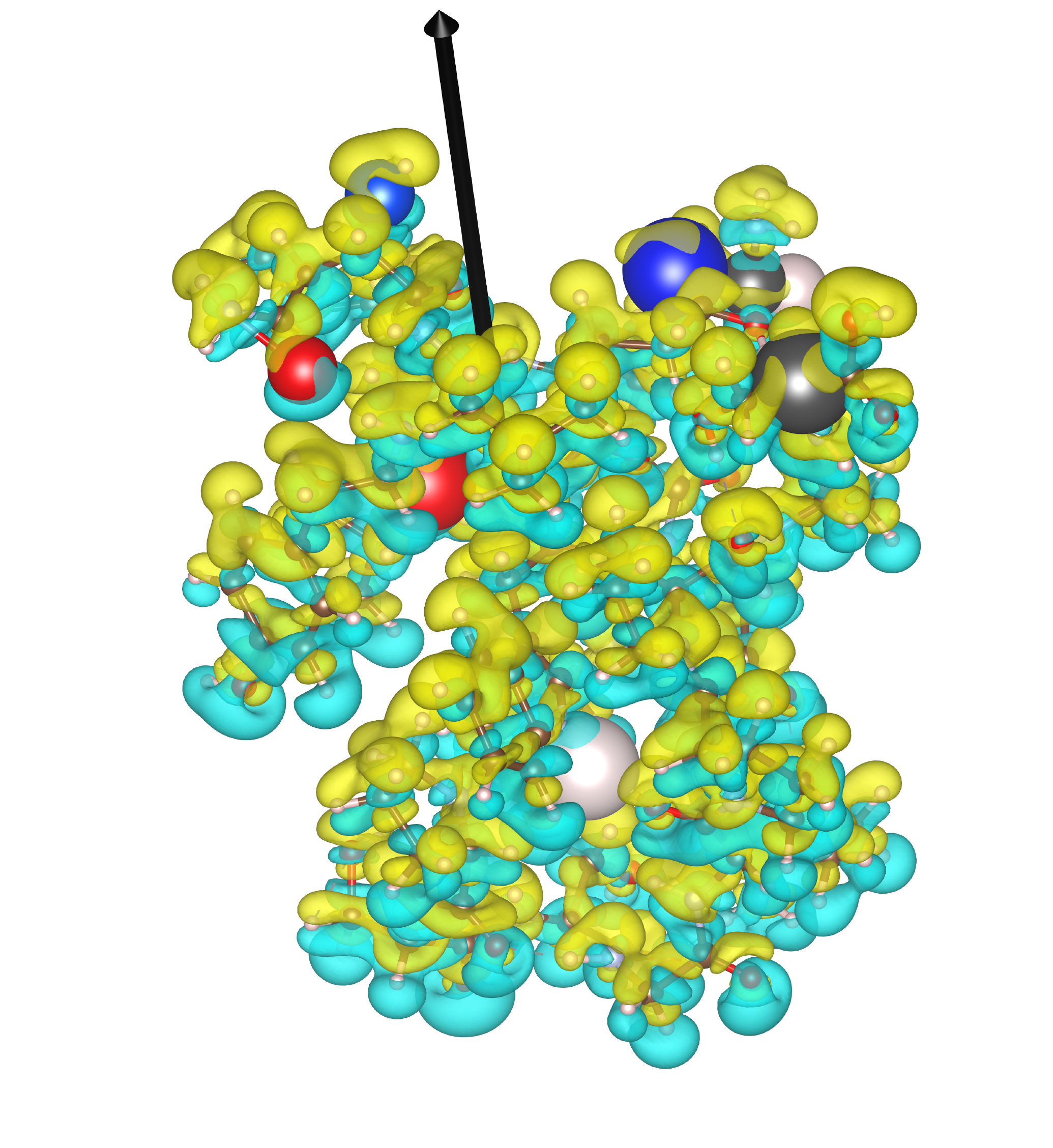}
\caption{\textit{Ab initio simulations} - The effect of the electric field on the electron distribution in the Trp-cage protein. Displayed is the difference in electron density between a protein un exposed to the field and a protein experiencing a 3 V/nm field. Blue electron density in the figure symbolizes a loss of electron density, and red an increase in electron density. Hydrogen atoms are white, Oxygen red, Nitrogen blue and Carbon black. Atoms whose correspond to a maximum decrease of the integrate electron number difference are depicted with atomic radii increased of a factor two; atoms whose correspond to a maximum increase of the integrate electron number difference are depicted with atomic radii enlarged of a factor three.  The isosurface level is set to 0.00226 electrons/\AA$^3$. The black arrow denotes the direction of the external electric field.}
\label{fig:eldens}
\end{figure}

\textit{Ab initio} calculations were performed on the Trp-cage protein. Given the extreme computational effort needed for quantum calculations, we limited  the extend of those simulations to comprise only the  electronic response, without nuclei dynamics. The main goal of these simulations is to have a quantitative estimation of the order of magnitude of the EF strength needed to break interatomic bonds, as well as validating the use of fixed charges in the classical MD simulations.\\
In Table~\ref{tab:bond_energies}, some of the most representative equilibrium bond forces within proteins are listed. These values were computed by dividing the tabulated bond energies by the tabulated equilibrium distances. We qualitatively assume that a bond among two atoms is broken when their distance is increased of 25\%  with respect to their equilibrium distance. We thus define a Bond Dissociation Force ($BDF_{thld}$) equal to  1 eV/ \AA\;as a reasonable lower estimation of the force sufficient to break a covalent bond. This value corresponds to $\approx$ 80\% of the force acting on two sulfur atoms in a  disulfide bond at  their  equilibrium distance  (Table~\ref{tab:bond_energies}).\\
We assume that the force acting on any atom $i$ in our simulations can be expressed as a sum of three terms, given as
\begin{equation}
    F_{i\;TOT}(E) = F(d_{CM}) + F(d_i) + F_{i\;field}(E).
\end{equation} 
 Here,  $F(d_{CM})$ is the force depending on the motion of the center of mass of the protein, $F(d_i)$ is the force due to the atomic vibrational state and $F_{field}(E)$ is the contribution to the force given by the interaction with the external EF. For each simulations at the different EF strength, we calculate the average relative force $\langle | F(E_j) | \rangle$ as
\begin{equation}
\label{eq:mean_force}
\langle | F(E_j) | \rangle = \frac{1}{N} \sum_{i}^{N}  | F_i(E_j) | -  | F_i(E=0) | = \frac{1}{N} \sum_{i}^{N} |F_{i\;field}(E)| .
\end{equation}
With this quantity we isolate the effects induced only by presence of the external EF from the total force at a given $E_j$. The contributions due to the center of mass motion and the particular vibrational state in which the atoms are frozen in are automatically removed. By comparing $\langle | F(E_j) | \rangle$ to $BDF_{thld}$ it is possible to infer an estimation of the field strength necessary to break a covalent bond. In Figure~\ref{fig:average_force} we display $\langle | F(E_j) | \rangle$ as function of the simulated EF. The plot shows that the EF strength needed to break atomic bonds is of the order of 45 V/nm. Moreover, one can notice that for  all the field strengths used in the classical MD (in the green inset of the graph), the value of the average force $ \langle | F(E_j) | \rangle$ is one order of magnitude lower than $BDF_{thld}$. This finding proves that the integrity of the protein topology is maintained in that field range  and the accuracy of classical MD is preserved.\\ 
In Figure~\ref{fig:eldens} the polarization induced by 3.0 V/nm field on the electron density is shown. At this field-strength, the majority of the polarization response is local to the atoms. Hence, not much charge is transferred across the molecule and the maximum charge increase and decrease in relation to the ground-state of a specific atom is relatively small. This validates the approach used in our classical MD simulations, where the ionic charge is fixed throughout the simulation. In Table~\ref{tab:max_charge}, we list the maximum increase and decrease in integrated charge on the atoms where the difference is largest. The corresponding position of the respective atoms are shown in Figure~\ref{fig:eldens}. The fact that these atoms are not residing at the edge of the protein also indicates that importance of the local polarization above de-localized charge transfer across the molecule, further corroborating the use of classical MD in this context. 

\begin{table}
\centering
\caption{\label{tab:bond_energies}  Covalent and hydrogen bond forces at the equilibrium of particular relevance in proteins. Force values are computed by dividing the tabulated energies by the tabulated equilibrium bond distances. Hydrogen bonds are indicated by $\dotsb$. }
\begin{tabular}{|c c | c c|}
\hline
\multicolumn{2}{c|} {\textbf{Covalent bonds}\cite{huheey2006inorganic}}  &  \multicolumn{2}{c} { \textbf{Hydrogen bonds}\cite{mcdaniel1983concepts}}  \\

Type & Force [eV/\AA] & Type & Force [eV/\AA] \\
\hline

\ch{C - N} & $\approx$ 2.0 & \ch{N-H} $\dotsb$ \ch{O} & $\approx$ 0.08\\
\ch{C - C} & $\approx$ 2.2 & \ch{C-H} $\dotsb$ \ch{N} & $\approx$ 0.12\\
\ch{C - S} & $\approx$ 1.4 & 
\ch{O-H} $\dotsb$ \ch{O} & $\approx$ 0.10\\
\ch{C - O} & $\approx$ 2.5 & \ch{C-H} $\dotsb$ \ch{O} & $\approx$ 0.20\\
\ch{S - S} & $\approx$ 1.3 & & \\
\hline
\end{tabular}

\end{table}

\begin{table}
\centering
\caption{\label{tab:max_charge} The maximum integrated electron number difference around each site for each atomic species. Units are in electronic charge $e$. The integrated difference is defined as `number of electrons on site X at field =0' - `number of electrons on site X with field=3 V/nm'. The corresponding atoms are highlighted in Figure~\ref{fig:eldens} by enlarging their radii of a factor two, for the maximum decrease, and a factor three, for the maximum increase. }
\begin{tabular}{|c | c | c|}
\hline
Atom type   &  Max charge increase &  Max charge decrease  \\
\hline
Hydrogen & 0.016 &  -0.018 \\
Carbon & 0.022 & -0.013\\
Nitrogen & 0.012 & -0.017\\
Oxygen & 0.010 & -0.014 \\
\hline

\end{tabular}
\end{table}

\subsection{Classical MD simulations: Orientation in time-dependent electric fields}
The way the protein responds to the time-dependent EF has been assessed by studying three different observables.
First of all, to assess the extent of protein orientation we define the {\textit{degree of orientation}  as
\begin{equation}
    \Theta = 1 - \cos(\theta),
\end{equation}
where $\theta$ is the angle between the EF and the total dipole moment of the molecule.
Thus, a fully aligned protein  expresses a value of $\Theta = 0$, whereas  $\Theta = 1$ corresponds to a perpendicular orientation when considering a protein at a particular moment in time.
$\Theta = 1$ is however also the expectation value for a randomly oriented protein, since parallel and anti-parallel orientations then are equally likely and the average $\cos(\theta)$ becomes $0$.

\begin{figure}[hbt!]
\centering
\includegraphics[width=0.65\linewidth]{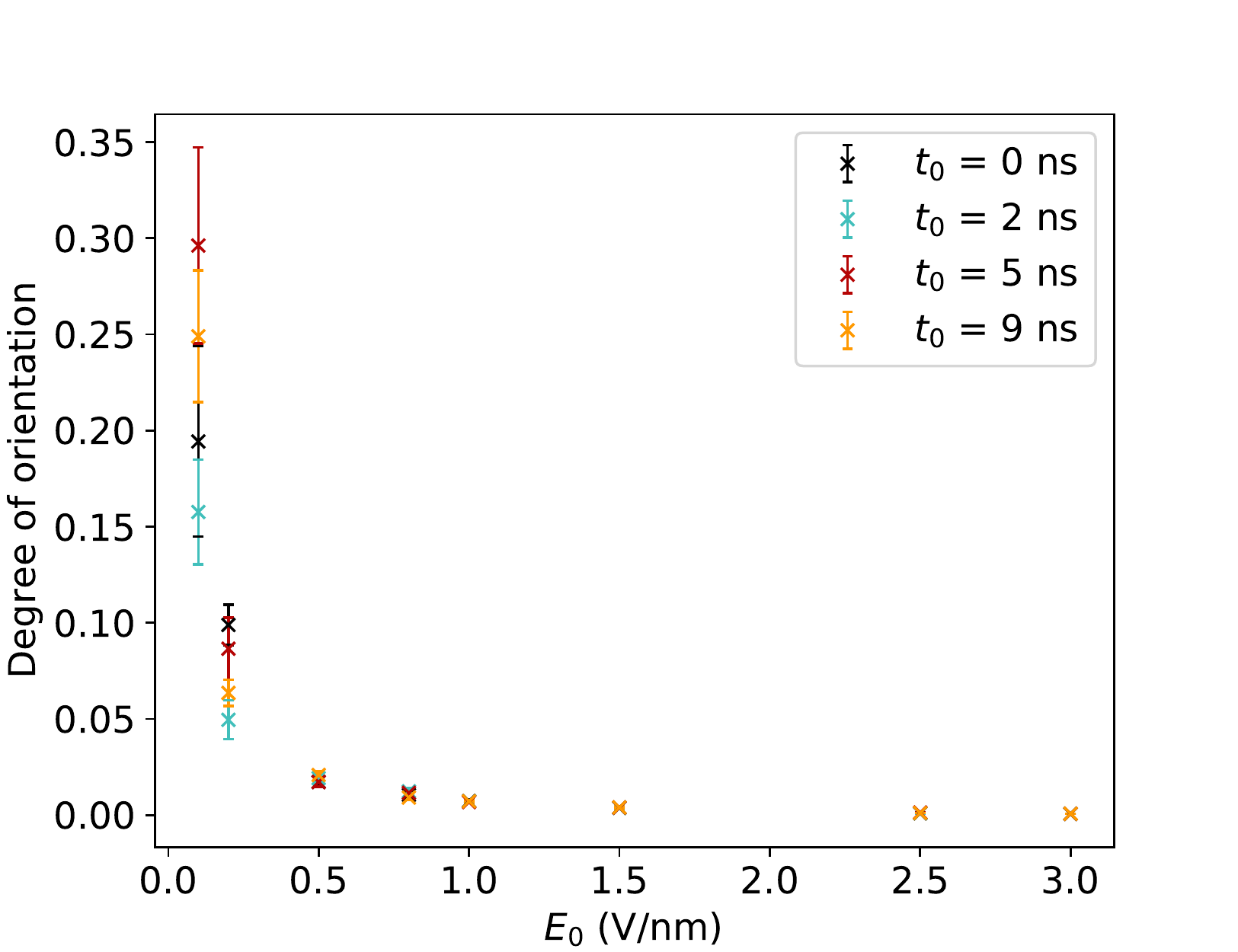}
\caption{\textit{Classical simulations} - An averaged degree of orientation as a function of the maximum field strength $E_0$ for different ramp-up times $t_0$. The different colors refer to the different field implementations (black - $t_0$= 0 ns, cyan - $t_0$= 2 ns, red - $t_0$= 5 ns and orange - $t_0$= 9 ns ).}
\label{fig:orientation_vs_E}
\end{figure}
In Figure~\ref{fig:orientation_vs_E}, we show how  $\Theta$  is depending  on EF strength $E_0$ and the  ramping time $t_0$.
Values in Figure~\ref{fig:orientation_vs_E} show the average degree of orientation over the last two nanoseconds of the ten independent simulations.
As one could expect, the stronger the field is, the more the molecule becomes orientated. In Figure~\ref{fig:proj_yz} (in Supporting Information section) the projection of the dipole moment of the plane perpendicular to the EF vector is depicted.  Here, $E_0$ strengths equal to 0.1 V/nm, for all four field implementations, are not strong enough to orient the protein in the simulation time we explored and the projection of the dipole moment is equally distributed in the plane. For $E_0$ equal to 0.2 V/nm, although there is not perfect alignment of the EF and the dipole of the molecule, the dipole moment distribution in not completely random. In particular, quite interestingly, the EF implementation with a ramping up time equal to 2 ns results in a better orientation respect to the other three EF implementations;  although not very focused, one can notice an highly populated region corresponding to a spread of $\pm$ 15 degrees with respect to perfect alignment. For fields values equal to 0.5 V/nm, the projection of the dipole moment in plane perpendicular to the field is focused in the $\pm$ 15 degrees region, expressing a  good alignment between the field vector and the ubiquitin dipole moment. For field strengths greater than or equal to 0.5 V/nm, the differences in the degree of orientation among the ramping times are not resolved within the errors (Figures~\ref{fig:orientation_vs_E}, ~\ref{fig:proj_yz_0.8_1.0_1.5} and~\ref{fig:proj_yz_2.5_3.0}}). At the end of the simulations, the  protein is orientated in a similar way, regardless of the field implementation. \\
\begin{figure}[hbt!]
\centering
\includegraphics[width=0.65\linewidth]{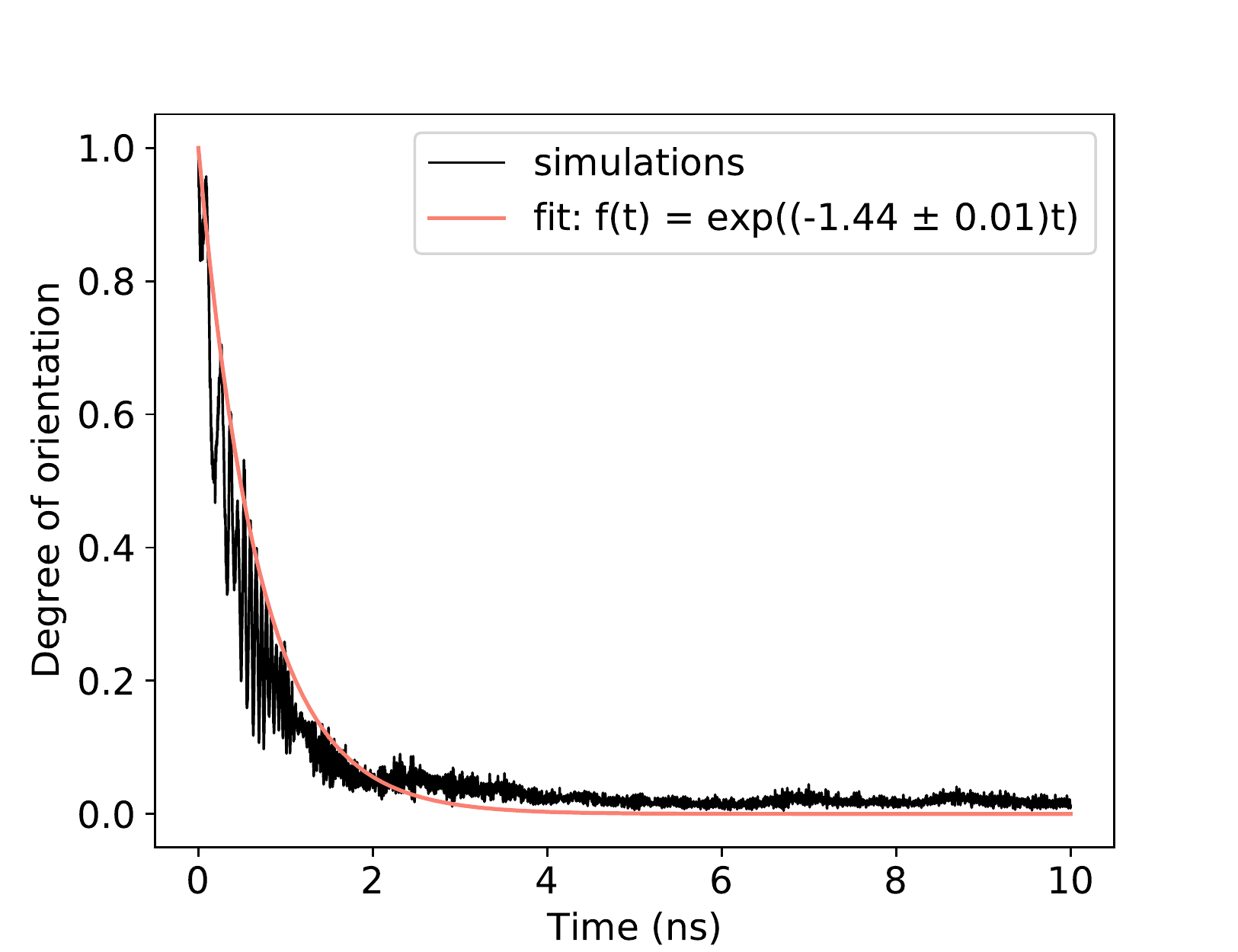}
\caption {\textit{Classical simulations} - The time evolution of degree of orientation averaged over ten independent runs for the parameters $E_0 = 0.5$ V/nm, $t_0=2$ ns (black line). The orange line is the result of fitting it with the function $f(t) = \exp(-kt)$.}
\label{fig:exp}
\end{figure}
The second observable we monitored in our simulations was the speed of orientation. 
Although the degree of orientation oscillates significantly, we can observe that there is a clear exponential decay. In Figure~\ref{fig:exp} an example of this trend for $E_{0}$=0.5 V/nm, $t_{0}$=2 ns is presented. Here,  the black line represents the evolution of $\overline{\Theta}$ over  time ($\overline{\Theta(t)}:= \langle \Theta(t)  \rangle _{replicas}$). The orange line represents a fit on $\overline{\Theta}(t)$, with  $f(t) = \exp(-kt)$.
We define $\tau$ as  the time required for the protein to lose 90\% of its initial orientation and hence to arrange parallel to the EF vector: 
\begin{equation}
    \tau = \frac{\ln(10)}{k}
\end{equation}

\begin{figure}[hbt!]
\centering
\includegraphics[width=0.65\linewidth]{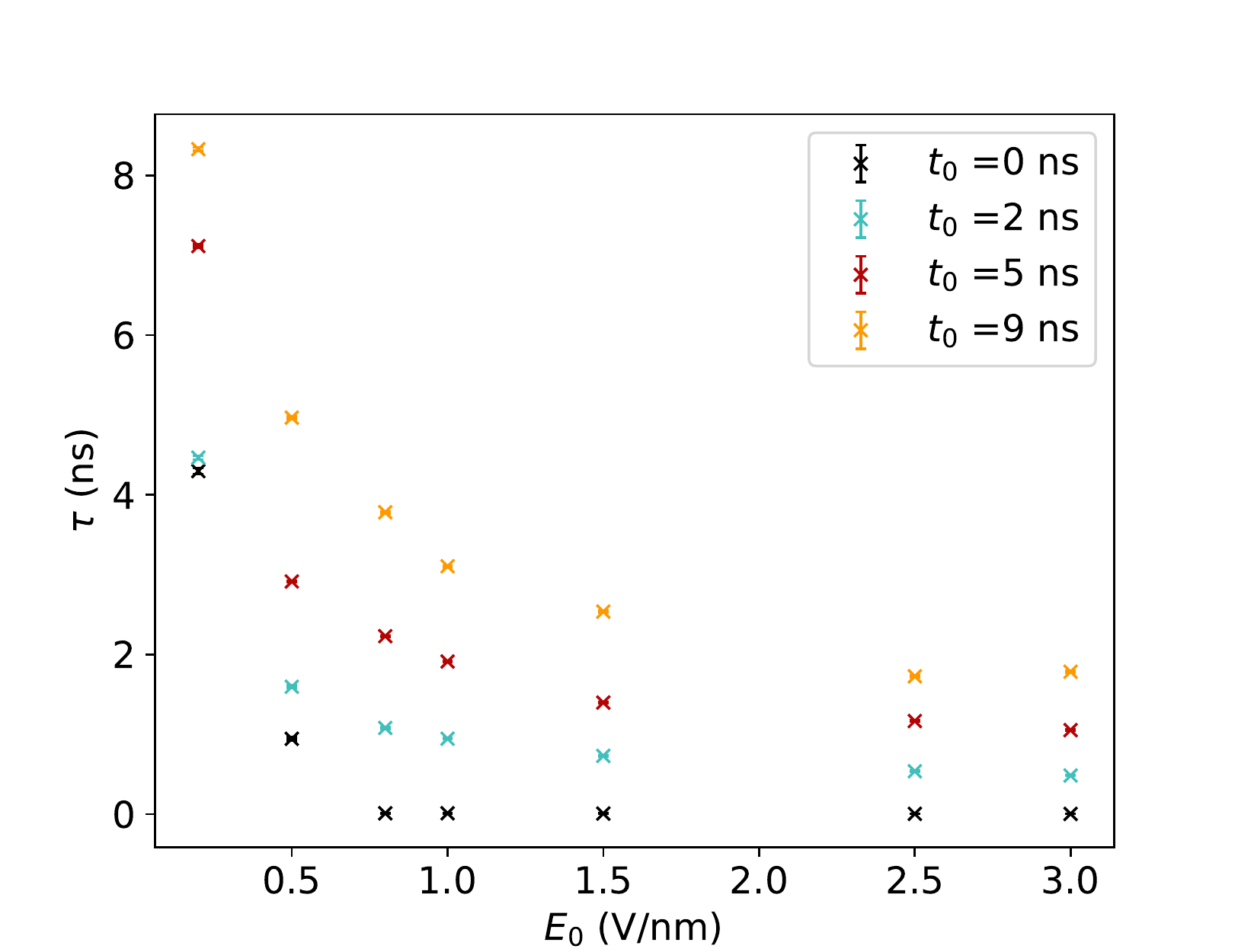}
\caption{\textit{Classical simulations} - The dependence of time $\tau$ (when the protein lost 90\% of initial orientation) on $E_0$ for different ramping time $t_0$. The colors scheme is the same one described in Figure~\ref{fig:orientation_vs_E}.}
\label{fig:t01}
\end{figure}
\begin{figure}[hbtp!]
\centering
\includegraphics[width=0.65\linewidth]{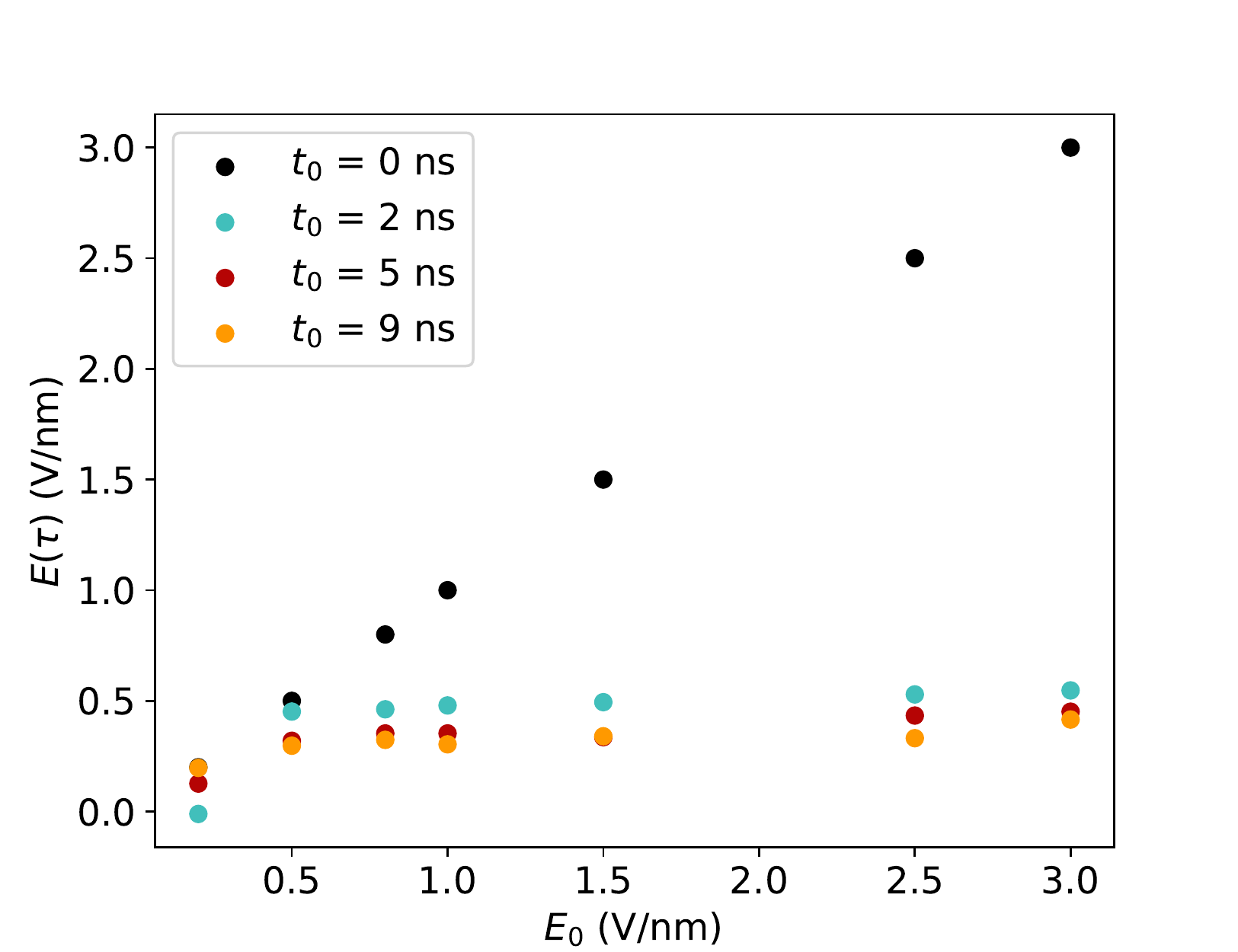}
\caption {\textit{Classical simulations} - EF strength at the time the protein is oriented as a function of the EF implementation. The colors scheme is the same described in Figure~\ref{fig:orientation_vs_E}.}
\label{fig:E(tau)}
\end{figure}

\begin{figure}[htbp!] 
\centering
\includegraphics[width=0.65\linewidth]{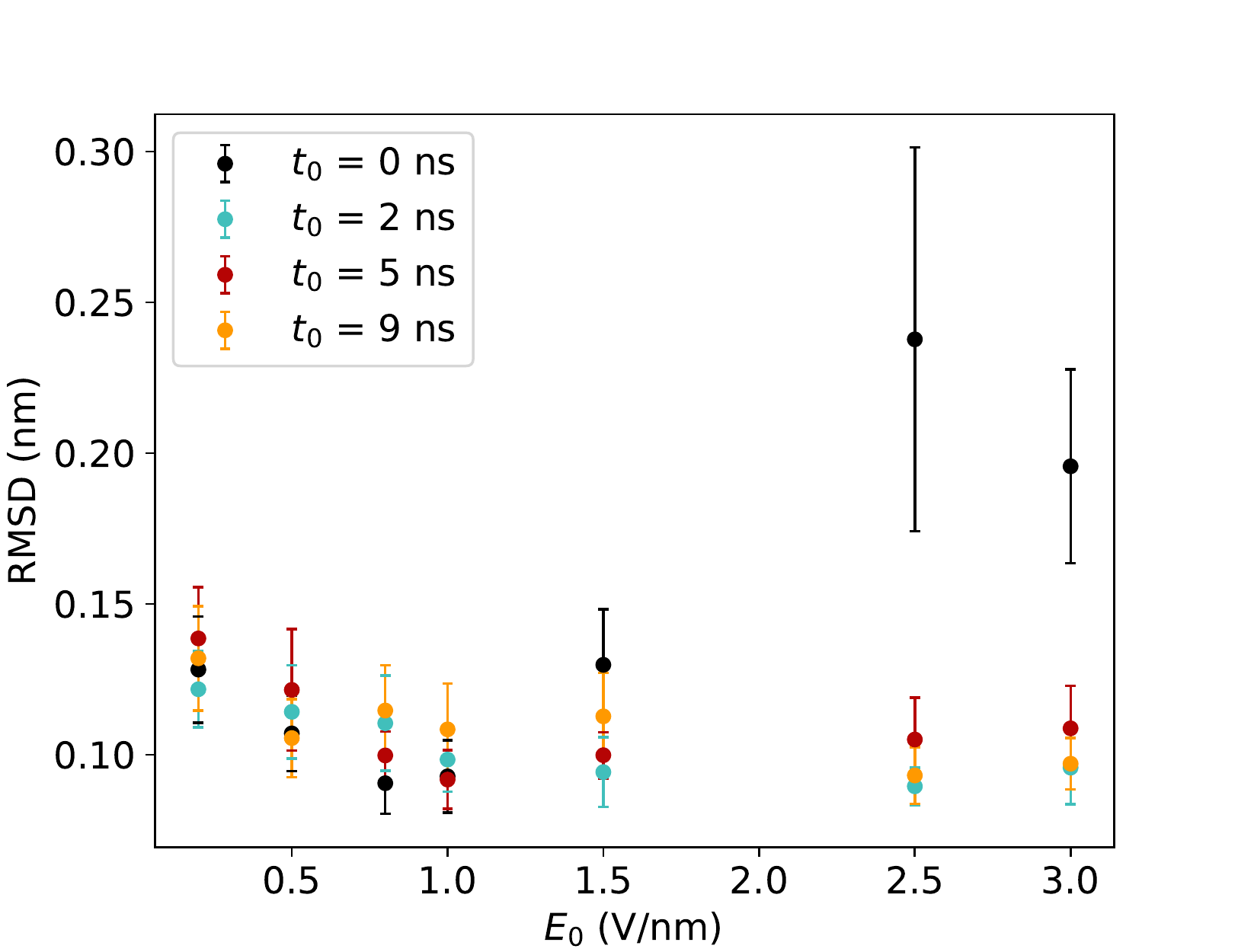}
 \caption{\textit{Classical simulations} -  The dependence of RMSD at time $t = \tau$ on $E_0$ for different ramping time $t_0$. The colors scheme is the same one described in Figure~\ref{fig:orientation_vs_E}.}
        \label{fig:rmsd_t0}
    \end{figure}

In Figure~\ref{fig:t01} we display the correlation between the orientation time $\tau$ for all the simulated field implementations $t_0$ and $E_0$. Evidently, a clear dependence between the rate of orientation and the ramping time can be observed: the longer the ramping time is, the more time is needed to orient the structure. Here, $\tau$ spans a range of values from 7.6 ns, in the case of the EF equal to 0.2 V/nm and $t_0$ equal to 9 ns, to 3 ps,  for the EF equal to 3.0 V/nm and $t_0$ equal to 0 ns.\\
The question naturally arises: \textit{is there any particular field strength able to orient the protein?} In order to answer this query we plot, in Figure \ref{fig:E(tau)}, $E(\tau)(E_0;t_0)$, namely the value of the EF at the time the protein is oriented as a function of the field implementations for the different simulations. Noteworthy, except for the trivial case of the constant field ($t_0$=0 ns), the field strength required to align the protein seems to be always of the order of 0.5 V/nm, independently of the final field strength of the simulation and the implementations. In other words, an EF strength of 0.5 V/nm is a necessary and sufficient condition to have a good alignment of ubiquitin.\\ 
 
\begin{figure}[htbp!]
\centering
\includegraphics[width=0.75\linewidth]{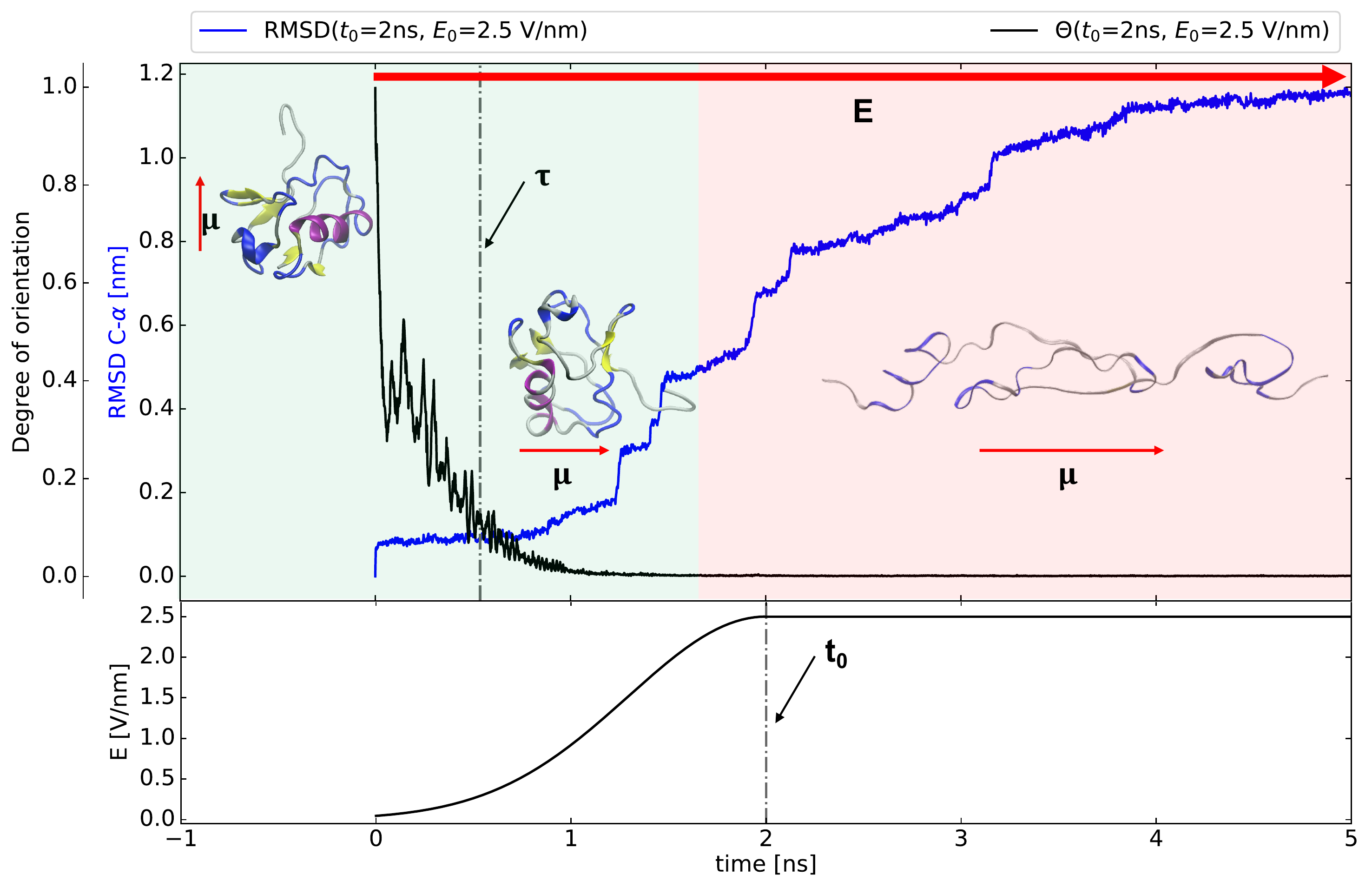}
\caption{\textit{Classical simulations} - Time evolution of the degree of orientation, RMSD and electric field function for $t_0$=2 ns and $E_0$= 2.5 V/nm simulations. Only the first 5 ns are shown. Each data point shows  the result of the averaged values of the 10 replicas. Error bars are not shown for simplicity. The green background represents the part of the simulation in which the ubiquitin structure is preserved (RMSD $\leq 0.5$ nm). On the contrary, the red background indicates the part of the simulation in which the protein structure is lost. The red arrow denoted with $E$ represents the direction of the external EF, with $\mu$ arrows we represent the direction of the protein dipole. In the insets, cartoon representations of the protein structure at the corresponding time are presented.
}
\label{fig:orient_bef_destr}
\end{figure}

The last important question we assessed regards the structure stability. It is of crucial importance to be aware of possible structural changes induced by the presence of an external EF. 
Root Mean Square Deviation (RMSD)\footnote{$RMSD(t_1-t_2)= \frac{1}{M}\sum_{i=1}{N} m_i \left|  \vec{r}_i(t_1)  - \vec{r}_i(t_2)    \right|^2$ where $M =\sum_{i=1}^{N} m_i$, $  \vec{r}_i(t)$ is the position of atom $i$ at time $t$ and $N$ is the total number of atoms of the system.} computed on C$\alpha$ atoms gives a measure of how much of the original structure is preserved. It is reasonable to define a structure to be preserved if the RMSD value is below 0.5 nm, while for RMSD values higher than this threshold, we can assume that the proteins initial structure is lost. \\
In Figure~\ref{fig:rmsd_t0}, the results of the RMSD($\tau$) calculations (namely the RMSD value at time $t = \tau$) as a function of $t_0$ and $E_0$  are displayed. We can observe that all ramping up times provide good conservation of the protein structures at the time it is orientated. This assumption is valid for all the values of the EF we tested. We can therefore conclude that, in all cases we simulated, the orientation happens before the structure is damaged, and as such "orientation before destruction".
This concept is visualized in Figure~\ref{fig:orient_bef_destr}.
    
\begin{figure}[htbp!]
\centering
\includegraphics[width=0.650\linewidth]{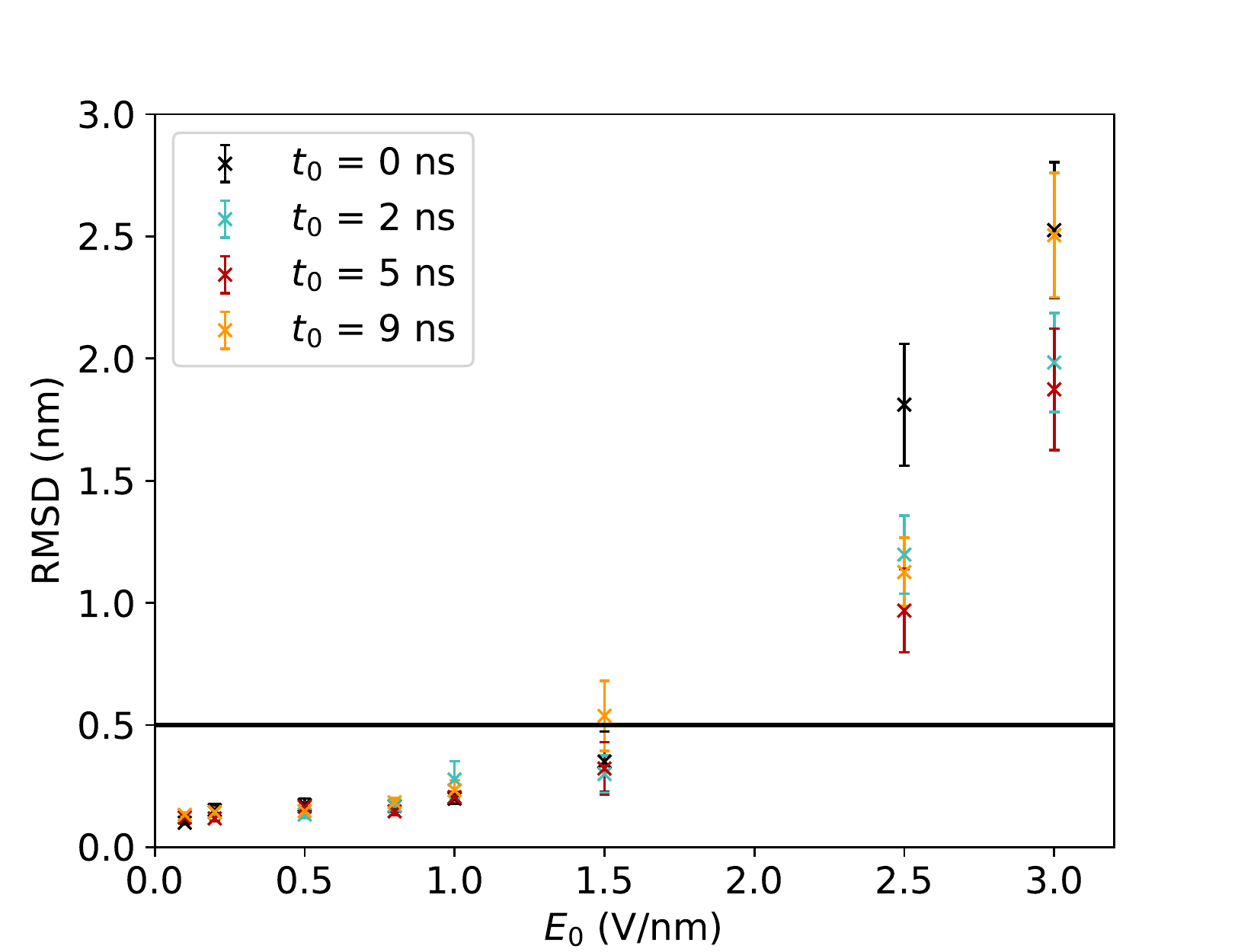}
\caption{\textit{Classical simulations} - Dependence of RMSD after 5ns of EF reached its  maximum value a function of $E_0$ and $t_0$. The colors scheme is the same one described in Figure~\ref{fig:orientation_vs_E}.}
        \label{fig:rmsd_5ns}    
\end{figure}

Moreover, it is interesting to measure how the protein structure evolves once it is fully immersed in an hypothetical experimental setup exerting an EF. In order to study this aspect, we evaluated the RMSD values of ubiquitin after 5 ns when the EF reached its maximum value. Namely we considered 5 ns from the beginning of the simulations for the case of $t_0$=0 ns, 7 ns for $t_0$=2 ns, 10 ns for $t_0$=5 ns, and lastly 14 ns for the case of $t_0$=9 ns. This describes accurately the possibility to acquire the image of the protein using the X-ray beam after the molecule travelled in the experimental device. The results of this analysis are presented in Figure~\ref{fig:rmsd_5ns}, where for each data point an average over the 10 replicas was computed. One can observe for EF lower than or equal to 1.5 V/nm, the protein structure is maintained for all the EF implementations. In particular for an EF lower than or equal to 0.8 V/nm, there are almost no differences among the four different EF implementations. On the contrary, EF field values greater than or equal to 2.5 V/nm express RMSD values above the 0.5 nm threshold for all the field implementations. Hence, it is possible to draw the conclusion that for these EF strengths the protein structure is lost.

\section{Discussion and conclusion}
 In the general framework of bio-imaging and in particular for the recent developed SPI technique, significant effort has been made to generate advanced algorithms aimed to solve the orientation problem for protein tertiary structure determination.  
All these approaches rely on the reconstruction of a 3D diffraction volume by assembling huge sets of diffraction patterns obtained from noisy, randomly oriented copies of a particle injected into an X-ray beam~\cite{bielecki2020perspectives}.
While such algorithms are very powerful, they cannot always converge to the correct solution, due to uncertainties and ambiguities in the data.
 The possibility to reduce some of the degrees of freedom of the problem by knowing the orientation during beam exposure is therefore of great interest. 
 
 One possible approach was proposed by Östlin \textit{et al.}~\cite{ostlin2018reproducibility}, where it was theoretically proven the opportunity of using an angular ion map to record information regarding the protein spatial orientation traceable from the sample Coulomb explosion. Here, instead of acquiring sparse orientation information from the experiment, we develop the approach of imposing the orientation onto the sample molecules, exploring the possibility to use an external, time-dependent EF to \textit{a prori} orient a molecule in the gas phase before its exposure to a X-ray beam. \\

First, we provided a qualitative indication on the EF strength required to alter considerably the chemical structures of the proteins. Given the impossibility of classical MD to study the electronic response to the external EF, we tackled this point by using DFT simulations to examine the Trp-cage mini-protein.
We studied the average atomic forces resulting from the interaction with the EF and we qualitatively estimated that the field strength needed to break interatomic bonds is of the order of 45 V/nm. This limit is considerably higher than what has been used in this context before~\cite{Marklund2017a,sinelnikova2020reproducibility}. Thus, we proved that the EF range used in the next part of the present study represents a safe windows of values where the interatomic forces acting on the protein atoms during EF exposure are lower than the force needed to break bonded interactions. In other words, classical MD approximation is still valid in our chosen EF range \footnote{It is noteworthy to recall that the possibility of breaking bonds is not modelled in classical MD in its standard parameterization, since the bonded interactions are given as input for all simulations.}.

We used classical MD simulations in vacuum to monitor the orientation of a protein as a response to a time-dependent EF. This study can be considered as a natural extension of our previous work~\cite{Marklund2017a}. In fact, the EF implemented here can be considered more  realistic with respect to the EF studied before, as it mimics the behavior of a molecule injected in an apparatus exerting an EF in greater detail. The injected molecule experiences a gradually increasing field strength, in part due to the experimental geometry, but also due to the non-square waveform of an applied electric field pulse.
Ubiquitin was exposed to an  external EF ranging from 0.1 V/nm to 3 V/nm. To reproduce different time-dependecies of the experienced EF, we tested three ramping up times ($t_0$= 2 ns, $t_0$= 5 ns and $t_0$= 9 ns) of the EF, namely the time required for the field to reach its maximum value. As a reference, we also studied the case of a static EF, referred to as $t_0$= 0 ns, that replicates the same EF implementation used in Marklund \textit{et al.}~\cite{Marklund2017a}. In order to collect enough statistics and to reproduce the intrinsic heterogeneity of SPI samples, we performed for each choice of $t_0$ and $E_0$ ten independent simulations that differ slightly in the initial protein structures. 
We found that for EF values greater and equal to 0.5 V/nm, regardless of the field implementation, desirable orientation of the protein was achieved. An EF of a strength equal to 0.1 V/nm results to be too weak to induce any orientation of ubiquitin (in the time scale  investigated herein). These findings are consistent with our previous work~\cite{Marklund2017a}. Quite interestingly, the orientation is not significantly dependent of the EF implementation. 
Given the fact that the initial structures of our replicas were oriented in the same way, we were in the position to study the velocity of orientation as a function of the EF implementation and field strength. The velocity of orientation was computed by performing an exponential fit on the degree of orientation versus time. We then defined $\tau$ to be the time required for the molecule to lose 90\% of its initial dipole orientation and to align to the EF. Studying $\tau$ as a function of $\{t_0, E_0\}$, we noticed, as expected, a linear dependence of the orientation time with $t_0$ and $E_0$. The  higher the field, the lower the time required to align the protein to the external field is. The lower $t_0$, the higher $\tau$.   
We found that for all time-dependent implementations of the EF we simulated, the EF strength required to orient the protein is of the order of 0.5 V/nm.

A careful monitoring of the protein structure progression was performed. We studied the evolution of C-$\alpha$ RMSD once the protein was exposed to the EF. We demonstrated that for all four EF implementations we simulated, the protein preserves its structure at the time it is oriented, $\tau$. For values of the field less then 1.5 V/nm, there is not relevant unfolding of the protein even for longer time scales. The unfolding of the protein happens after the orientation - `orientation before destruction'. 

\section*{Conflicts of Interest} The authors declare that they have no conflicts of interest with the contents of this article.

\section*{Author Contributions}
C.C., E.G.M. and E.D.S. devised the project. C.C. with help from O.G. performed {\it ab initio} simulations. H.A.,  T.M. and A.S. carried out classical simulations.  A.S. and E.D.S. analyzed the data. E.D.S. with the help from  C.C., O.G., E.G.M. and A.S. wrote the article. 

\section*{Acknowledgments}
This work is a part of the MS SPIDOC project funded by the European Union’s Horizon 2020 FET-OPEN research and innovation program (Grant agreement No. 801406) for support of ongoing photo activation IM-MS activities. A.S. acknowledges funding from the Knut and Alice Wallenberg foundation through the Wallenberg Academy Fellow grant of J. Nilsson. C.C. acknowledges the Swedish Research Council (2018-00740) and the Helmholtz Association through the Center for Free-Electron Laser Science at DESY. 
E.D.S. gratefully acknowledges a postdoctoral fellowship from Carl Trygger foundation.\\
We acknowledge the use of Uppsala Multidisciplinary Center for Advanced Computational Science (Uppmax projects number SNIC 2020/15-67, SNIC 2019/8-314 and SNIC 2019/30-47) provided by SNIC. \\
We thank Maxim Brodmerkel for comments that greatly improved the manuscript.

\section*{Data availability}
The data that support the findings of this study are available from the corresponding author upon reasonable request.

\bibliography{main}

\input{SI.tex}

\end{document}

%% file: SI.tex
\clearpage
\renewcommand\thefigure{S\arabic{figure}} 
\renewcommand{\thetable}{S\arabic{table}}  
\setcounter{figure}{0}
\setcounter{table}{0}
\section*{Supporting information}

\label{section:SI}

\subsection*{\textit{ab initio} MD simulations}

\begin{table}[H]
\centering
\caption{\label{tab:dip_mom_max_force}  $\langle | F(E) | \rangle $ as a function of the electric field strength. Values are given in eV/\AA\; units.}
\begin{tabular}{|c|c|}
\hline
\textbf{Electric field}   & $\langle | F(E) | \rangle$ \\ 
\hline
0.0 & 0.000   \\
0.5 & 0.010  \\
1.0 & 0.020  \\
1.5 & 0.030  \\
2.0 & 0.041  \\
2.5 & 0.051  \\
3.0 & 0.061  \\
4.0 & 0.080  \\
5.0 & 0.097  \\
6.0 & 0.112  \\
7.0 & 0.127  \\
8.0 & 0.141 \\
9.0 & 0.155  \\
10.0 & 0.166  \\
20.0 & 0.311 \\
25.0 & 0.411  \\
30.0 & 0.523  \\
35.0 & 0.652 \\
40.0 & 0.810 \\
45.0 & 0.984  \\
50.0 & 1.179  \\
\hline
\end{tabular}
\end{table}
\subsection*{Classical MD simulations}
\begin{figure*}[hbt!]
\centering
\includegraphics[width=0.8\linewidth]{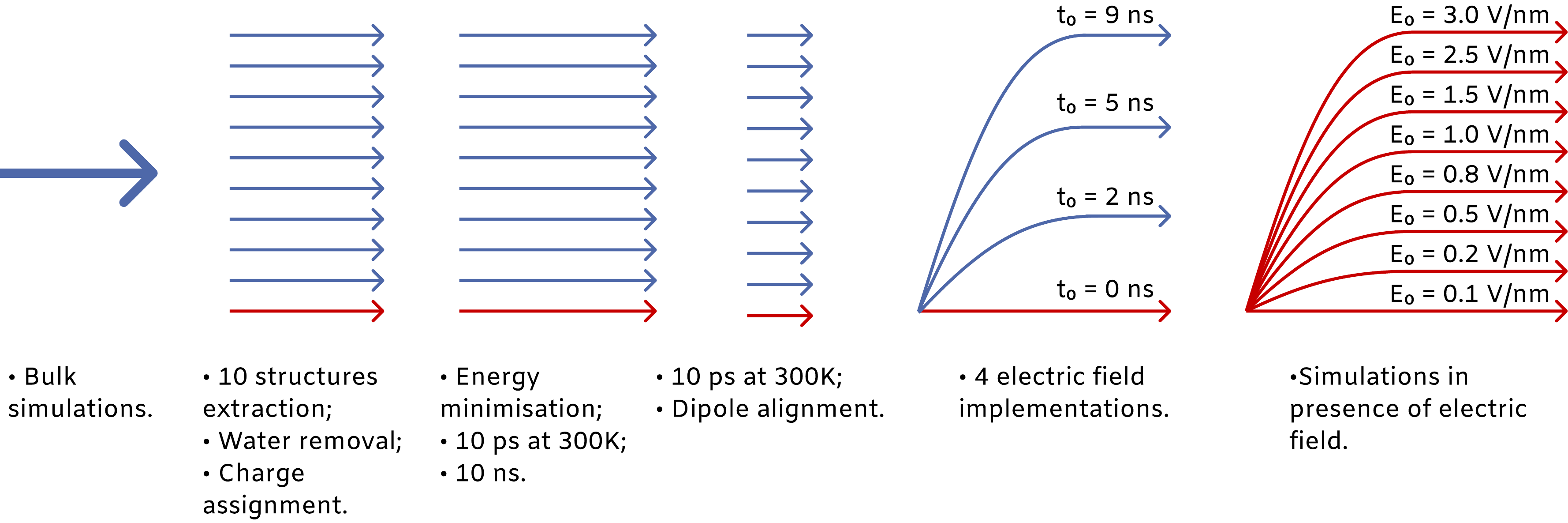}
\caption{The schematic representation of performed classical MD simulations.}
\label{fig:simScheme}
\end{figure*}

\begin{figure*}[hbt!]
\centering
\includegraphics[width=1\linewidth]{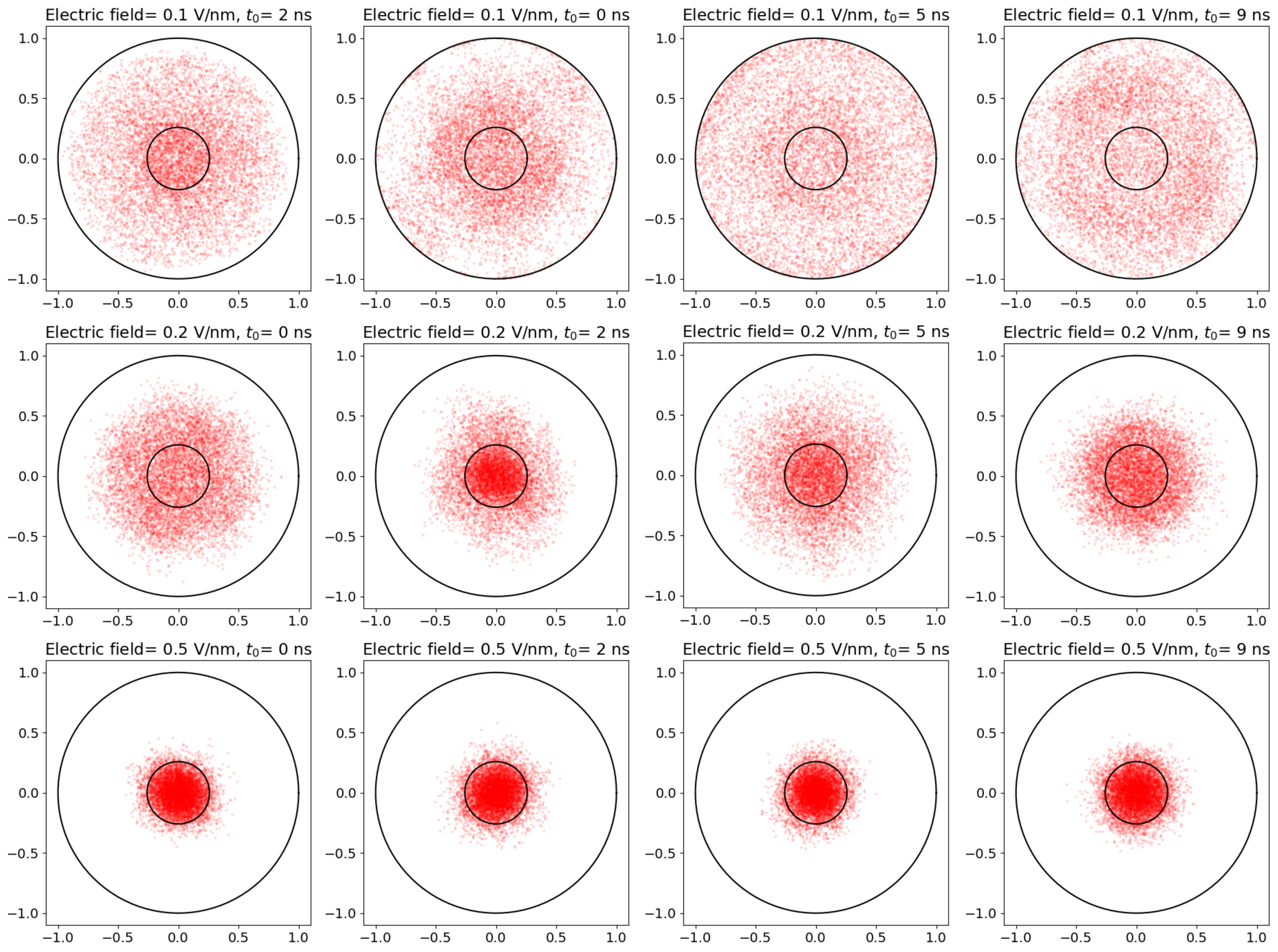}
\caption{Scatter plot of the projection of the dipole moment on the yz plane for the simulations with electric field strength of 0.1 V/nm (first row), 0.2 V/nm (middle row) and 0.5 V/nm (lower row) . Each point of every panel refers to a frame taken in the last 2 ns of simulations for all the ten replicas. Each panel contains 8000 points (the frames are taken every 2.5 ps of simulations). The circle drawn in the center of the panels represent an uncertainty of $\pm$ 15 degrees respect to a perfect alignment.}
\label{fig:proj_yz}
\end{figure*}

\begin{figure*}[hbt!]
\centering
\includegraphics[width=1\linewidth]{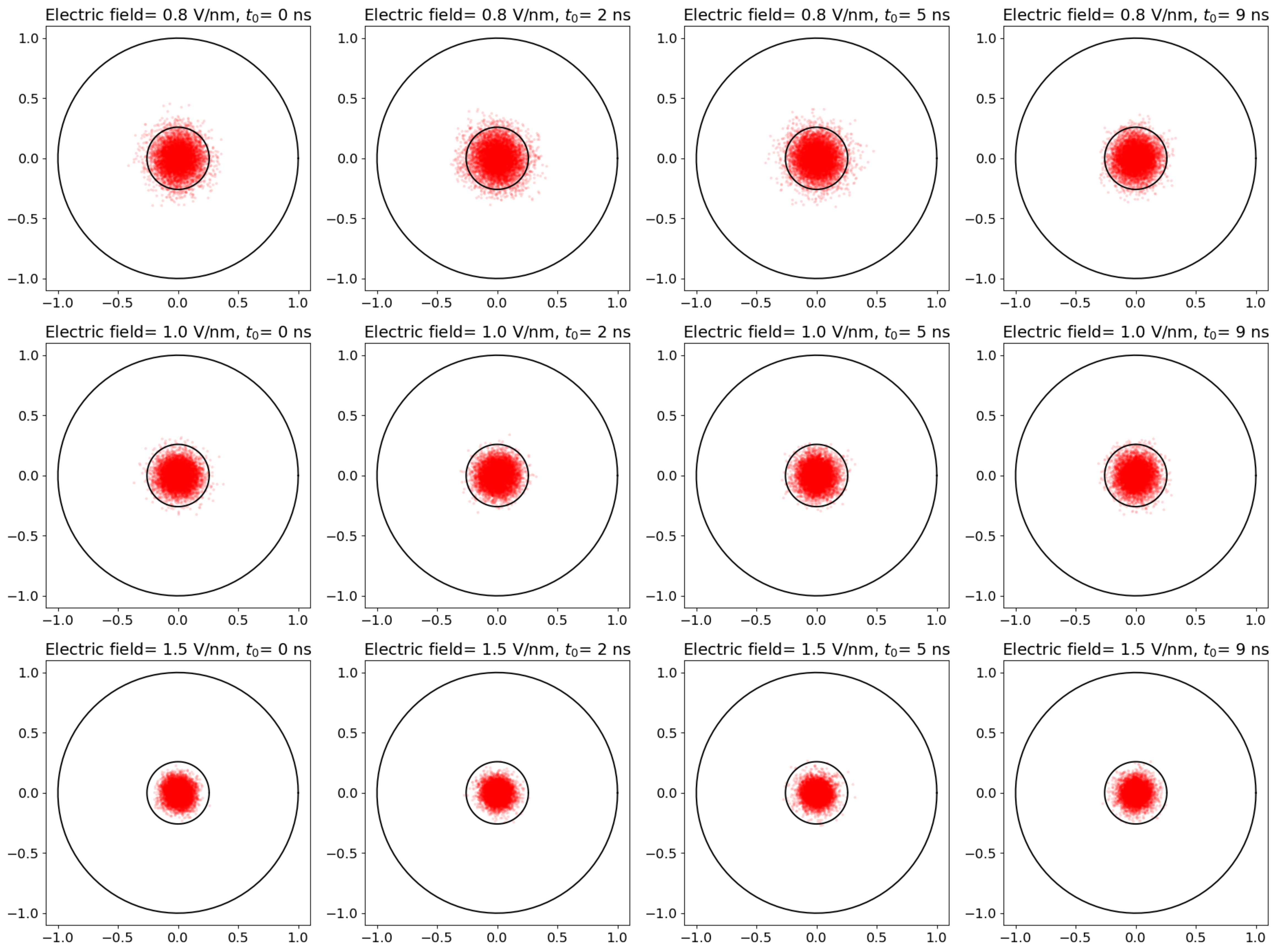}
\caption{Scatter plot of the projection of the dipole moment on the yz plane for the simulations with electric field strength of 0.8 V/nm (first row), 1.0 V/nm (middle row) and 1.5 V/nm (lower row) . Each point of every panel refers to a frame taken in the last 2 ns of simulations for all the ten replicas. Each panel contains 8000 points (the frames are taken every 2.5 ps of simulations). The circle drawn in the center of the panels represent an uncertainty of $\pm$ 15 degrees respect to a perfect alignment.}
\label{fig:proj_yz_0.8_1.0_1.5}
\end{figure*}

\begin{figure*}[hbt!]
\centering
\includegraphics[width=1\linewidth]{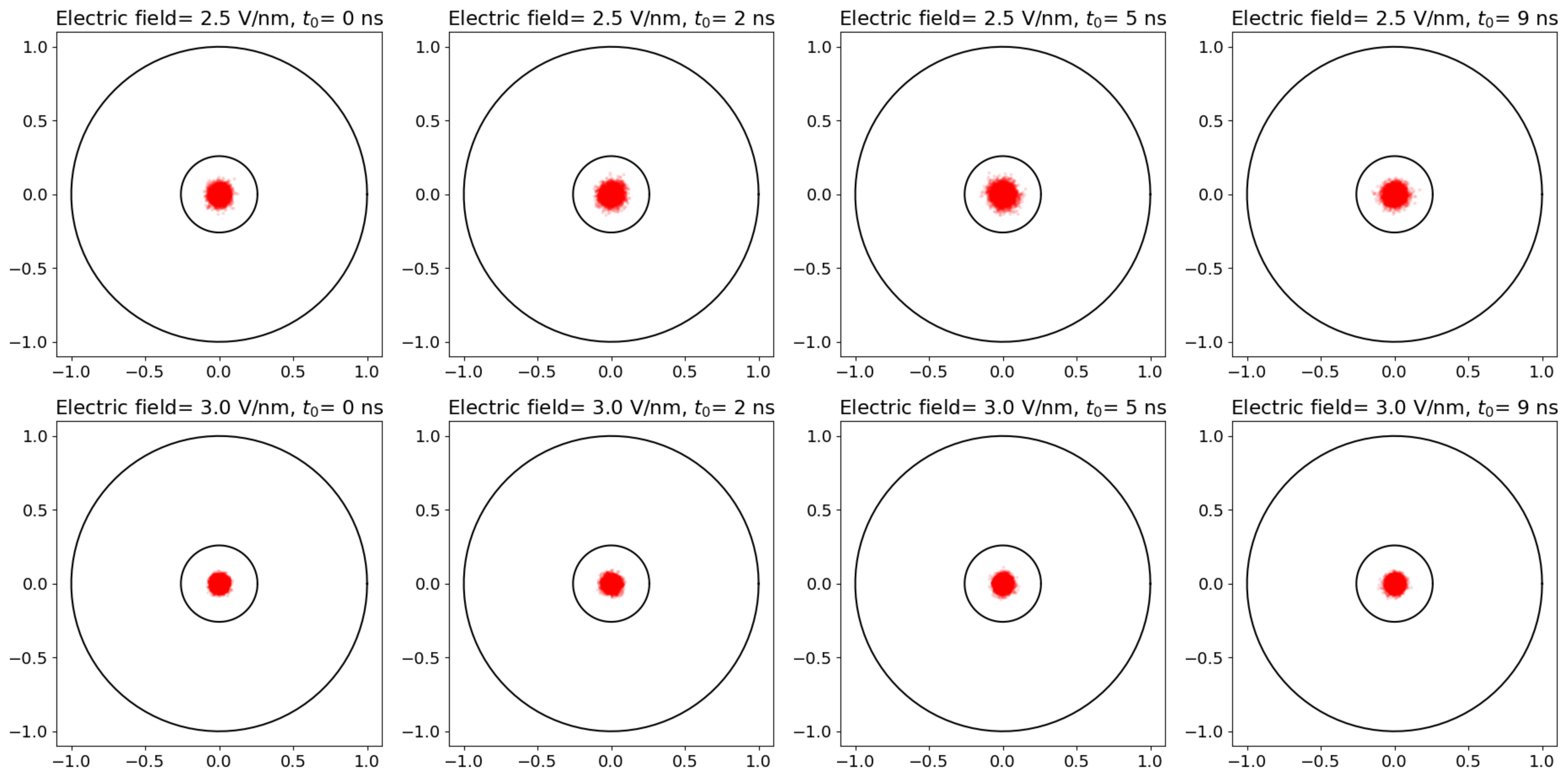}
\caption{Scatter plot of the projection of the dipole moment on the yz plane for the simulations with electric field strength of 2.5 V/nm (first row) and 3.0 V/nm (lower row) . Each point of every panel refers to a frame taken in the last 2 ns of simulations for all the ten replicas. Each panel contains 8000 points (the frames are taken every 2.5 ps of simulations). The circle drawn in the center of the panels represent an uncertainty of $\pm$ 15 degrees respect to a perfect alignment.}
\label{fig:proj_yz_2.5_3.0}
\end{figure*}